\documentclass[conference]{IEEEtran}
\IEEEoverridecommandlockouts
\usepackage{cite}
\usepackage{amsmath,amssymb,amsfonts}
\usepackage{algorithmic}
\usepackage{graphicx}
\usepackage{textcomp}
\usepackage{xcolor}
\usepackage{float}
\usepackage[nolist,nohyperlinks]{acronym}
\usepackage[hyphens]{url}
\usepackage[hidelinks]{hyperref}
\usepackage[exponent-product = \cdot,separate-uncertainty=true]{siunitx}
\def\BibTeX{{\rm B\kern-.05em{\sc i\kern-.025em b}\kern-.08em
    T\kern-.1667em\lower.7ex\hbox{E}\kern-.125emX}}

\usepackage{color}
\definecolor{green}{rgb}{0.,0.5,0.5}
\definecolor{blue}{rgb}{0.75,0.75,0}
\definecolor{magenta}{rgb}{0.79, 0.08, 0.48}
\definecolor{blue}{rgb}{1.0,0.33,0}

\newcommand{\printlengths}{
\typeout{TEXTWIDTH=\the\textwidth}
\typeout{COLUMNWIDTH=\the\columnwidth}
}
\AtBeginDocument{\printlengths}

\begin{document}
\begin{acronym}
    \acro{MMR}{magneto-mechanical resonator}
    \acro{MSE}{mean squared error}
    \acro{TxRx}{transmit-receive}
    \acro{CI}{confidence interval}
    \acro{GT}{ground-truth}
\end{acronym}

\title{Data-Driven Localization of Magneto-Mechanical Resonators from Time Signals
\thanks{This project is funded by the Deutsche Forschungsgemeinschaft (DFG, German Research Foundation) - SFB 1615 - 503850735 and 555456057.}
}


\author{\IEEEauthorblockN{Artyom Tsanda\IEEEauthorrefmark{1}\IEEEauthorrefmark{2},
Konrad Scheffler\IEEEauthorrefmark{1}\IEEEauthorrefmark{2},
Sarah Reiss\IEEEauthorrefmark{1}\IEEEauthorrefmark{2},
Jonas Faltinath\IEEEauthorrefmark{1}\IEEEauthorrefmark{2}, \\
Janik Bach\IEEEauthorrefmark{1}, 
Florian Thieben\IEEEauthorrefmark{1}\IEEEauthorrefmark{2}\IEEEauthorrefmark{3},
Pascal Stagge\IEEEauthorrefmark{3}, 
Fabian Mohn\IEEEauthorrefmark{1}\IEEEauthorrefmark{2}, and
Tobias Knopp\IEEEauthorrefmark{1}\IEEEauthorrefmark{2}\IEEEauthorrefmark{3}
}
\IEEEauthorblockA{
\IEEEauthorrefmark{1}Institute for Biomedical Imaging, Hamburg University of Technology, Hamburg, Germany\\
\IEEEauthorrefmark{2}Section for Biomedical Imaging, University Medical Center Hamburg-Eppendorf, Hamburg, Germany\\
\IEEEauthorrefmark{3}Fraunhofer Research Institution for Individualized Medical Technology and Engineering IMTE, Lübeck, Germany \\
Email: artyom.tsanda@tuhh.de
}}

\maketitle
\begin{abstract}
    Magneto-mechanical resonators (MMRs) are a recently introduced sensor technology for tracking and sensing applications
    that enables effective miniaturization down to the millimeter regime. 
    However, estimating sensor parameters such as position from raw signals remains computationally challenging,
    because it requires solving a nonlinear inverse problem with computationally demanding objective function evaluations. 
    In this work, we show how a deep learning method can accelerate this process by directly learning the inverse mapping from the measured signals.
    We consider the localization problem as an example and show that 
    a convolutional-transformer regressor trained on simulated data with recorded noise 
    can generalize to experimental data. 
    The method achieves a median Euclidean distance error of 7~mm, which is comparable to the size of the sensor and corresponds to 6\% of the size of a single coil. 
    These results suggest that deep learning is a promising approach for rapid estimation of MMR parameters.
\end{abstract}


\section{Introduction}

Tracking medical devices is an important research direction for biomedical applications, 
such as the localization of medical tools in minimally invasive surgeries,
where optical and electromagnetic tracking systems are commonly used~\cite{8822749}.
Recently, a novel tracking and sensing technology based on a \ac{MMR} has been proposed~\cite{Gleich}, 
which allows more effective miniaturization than electromagnetic sensors by providing higher signal-to-noise ratio.

\begin{figure}[tbp]
    \centering
    \includegraphics[width=\columnwidth]{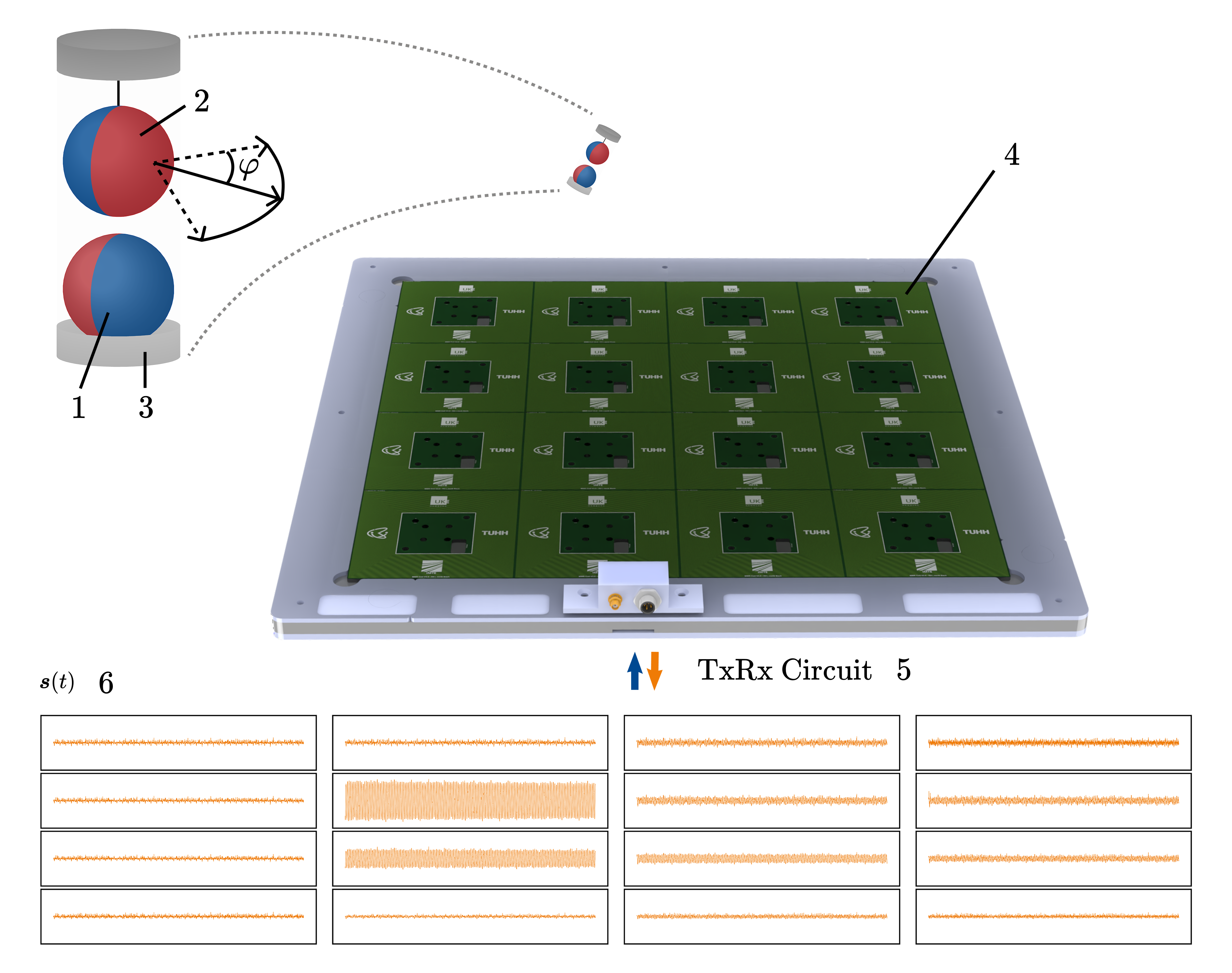}
    \caption{
        Schematic overview of the experimental setup with the magneto-mechanical resonator
        (stator (1), rotor (2), housing (3)), the inductive coil array (4), and the
        transmit-receive circuit for signal generation and acquisition (5).
        An exemplary 16-channel signal (6) is shown below.
    }
    \label{fig:mmr-setup}
\end{figure}

The considered \ac{MMR} consists of two magnets aligned antiparallel, where one magnet is fixed to the housing (stator) and 
the other is attached via a thin filament that allows torsional oscillations (rotor) (see Fig.~\ref{fig:mmr-setup}).
The oscillating magnetic moment of the rotor induces voltage signals in the receive coils, 
which allow inference of the position and orientation of the sensor. 
Additionally, sensing parameters such as temperature~\cite{11331063}, pressure~\cite{Merbach_2025}, and pH~\cite{Timm2026_DPG} can be encoded into the sensor's frequency.

Estimation of sensor parameters from measured signals is a challenging, computationally intensive problem, 
as it involves solving a nonlinear inverse problem with the objective function 
relying on numerical ODE solvers and numerical integration of the Biot-Savart law~\cite{reiss_parameter_2026}.
Consequently, fast estimation methods are required to achieve the computation times needed for real-time applications such as catheter tracking in interventional medical procedures~\cite{Gleich}. 

Deep learning has shown great potential for solving inverse problems 
by approximating the inverse mapping from the measurements to the parameters of interest~\cite{Choudhary_Sadak_Gerena_Haliyo_2025}.
In this work, we introduce a proof-of-concept deep learning approach for the localization of an \ac{MMR}, 
where the position is estimated directly from the time signals measured by an inductive coil array. Source code~\cite{tsanda_code_supplement} and data~\cite{tsanda_data_supplement} used in this study are publicly available to ensure reproducibility.

\section{Methods}

\subsection{Signal Model}

Due to the attractive force between the magnets, the rotor aligns antiparallel in stable equilibrium. 
When displaced from this orientation, the rotor starts to oscillate and returns to the equilibrium alignment in the absence of external magnetic forces.
Neglecting the impact of the gravitational force and assuming the rotor has only a torsional degree of freedom, the free oscillation of 
the rotor's magnetic moment $\pmb{m}: [0, T] \rightarrow \mathbb{R}^3$, where $T$ is the measurement duration, 
can be described by the following second-order ordinary differential equation~\cite{Gleich}:
\begin{equation} \label{Eq:DeflAngleODE}
\begin{cases}
\ddot{\varphi} + \frac{\omega_\text{nat}}{Q} \dot{\varphi} + \omega_\text{nat}^2 \sin{\varphi} = 0 \\
\varphi(0) = \varphi_\text{max} \cos(\psi_\text{start}) \\
\dot{\varphi}(0) = -\dot{\varphi}_\text{max} \sin(\psi_\text{start}),
\end{cases}
\end{equation}
where 
$\varphi: [0, T] \rightarrow (-\frac{\pi}{2}, \frac{\pi}{2})$ is the deflection angle of the rotor from its equilibrium orientation, 
$\omega_\text{nat} \in \mathbb{R}_0^+$ is the natural frequency of the resonator, 
$Q \in \mathbb{R}_0^+$ is the quality factor, 
$\varphi_\text{max} \in [0, \frac{\pi}{2})$ and 
$\dot{\varphi}_\text{max} \in \mathbb{R}$ are the maximum deflection angle and angular velocity, respectively, 
and $\psi_\text{start} \in [0, 2\pi)$ is the initial phase. 

According to Faraday's law of induction, the voltage signals $\pmb{s}: [0,T] \rightarrow \mathbb{R}^L$ induced in the $L$ receive coils can be expressed as
\begin{equation} \label{Eq:SignalModel}
\pmb{s}(t;\pmb{r},\pmb{R}) = -\pmb{P}(\pmb{r}) \pmb{R}\, \partial_t \pmb{m}(t) + \pmb{n}(t),
\end{equation}
where $\pmb{P}: \mathbb{R}^3 \rightarrow \mathbb{R}^{L \times 3}$ is the coils' sensitivity map, 
$\pmb{r} \in \mathbb{R}^3$ and $\pmb{R} \in SO(3)$ are the rotor's position and orientation, respectively, 
$\partial_t \pmb{m}(t)$ is the time derivative of the magnetic moment of the rotor, and $\pmb{n}: [0,T] \rightarrow \mathbb{R}^L$ is the measurement noise.

\subsection{Data Generation}

Employing the signal model described in \eqref{Eq:SignalModel} and \eqref{Eq:DeflAngleODE}, 
we simulate training and validation datasets of sizes \num{100000} and \num{10000}, respectively. 
For each signal, the simulation parameters $\mathcal{P} := \{ \pmb{r}, \pmb{R}, \omega_\text{nat}, Q, \psi_\text{start}, \varphi_\text{max} \}$ are sampled randomly, covering the expected range of values 
in the real-world experiment. The corresponding distributions are summarized in Table~\ref{table:simulation_parameters}, where $\mathcal{U}(a, b)$ denotes the uniform distribution on the interval $[a, b]$.
The sampling frequency and the number of sampled points $N$ remain fixed for all samples and are set 
to \SI{61035}{Hz} and \num{53700}, respectively.
We solve the resulting initial value problem in~\eqref{Eq:DeflAngleODE} 
using the Tsitouras 5/4 Runge-Kutta method~\cite{tsitouras2011runge}. 
For each position, the sensitivity $\pmb{P}(\pmb{r})$ is inferred by numerically integrating the Biot-Savart law 
for the given geometry of the receive coil array used in the experiment. 
Because the system is operated in an unshielded environment, the noise $\pmb{n}(t)$ is difficult to model with a predefined distribution. 
Therefore, we record noise samples from the real system and add them to the simulated signals. 

\begin{table}[bp]
\caption{Simulation parameters and their respective distributions.}
\begin{center}
\begin{tabular}{|l c|c|}
\hline
\multicolumn{2}{|c|}{\textbf{Parameter}} & \textbf{Distribution} \\
\hline
Position,\,\unit{m}  & $\pmb{r}$ & $\mathcal{U}([-0.2,0.2]^2 \times [0.005,0.2])$ \\
Orientation & $\pmb{R}$ & $\mathcal{U}(SO(3))$ \\
Angular natural freq.,\,\unit{Hz} & $\omega_\text{nat}$ & $2\pi \cdot \mathcal{U}(177,\, 186)$ \\
Quality factor,\,\unit{s} & $Q$ & $\mathcal{U}(3200,\,7200)$ \\
Initial phase,\,\unit{rad} & $\psi_\text{start}$ & $\mathcal{U}(0,\,2\pi)$ \\
Max. def. angle,\,\unit{rad} & $\varphi_\text{max}$ & $\mathcal{U}(0.2,\,0.4)$ \\
\hline
\end{tabular}
\label{table:simulation_parameters}
\end{center}
\end{table}

\subsection{Network Architecture and Training}

We introduce a deep learning model 
$f_{\pmb{\theta}}: \mathbb{R}^{L \times N} \rightarrow \mathbb{R}^3$, parameterized by $\pmb{\theta} \in \mathbb{R}^P$, where $P$ is the number of parameters. The model maps the signals $\pmb{S} \in \mathbb{R}^{L \times N}$ 
to the position of the rotor $\pmb{r}$. 
The parameters $\pmb{\theta}$ are optimized by minimizing the \ac{MSE} loss function defined as:
\begin{equation}
    \min_{\pmb{\theta}} \mathbb{E}_{(\pmb{S}, \pmb{r}) \sim \mathcal{D}} \| f_{\pmb{\theta}}(\pmb{S}) - \pmb{r} \|^2,
\end{equation}
where $\mathcal{D}$ is the distribution of the training data.

The proposed architecture comprises three blocks: 
a convolutional feature extractor, a transformer encoder layer for temporal dependencies, and a linear output layer for regression.
The convolutional block consists of two 1D convolutions with kernel size $5$, which expand the channel dimension from $L$ to $E/2$ to $E$, where $E = 128$ is the embedding dimension.
Each convolution is followed by batch normalization and a GELU nonlinearity~\cite{hendrycks2017bridging}.
The transformer encoder layer follows the original architecture~\cite{Vaswani20175999} with layer normalization replaced by batch normalization.
The multi-head attention has $8$ heads, and the feed-forward network has an expansion factor of $4$.
Finally, global average pooling over the temporal dimension aggregates the sequence representation, 
and a linear output layer maps the pooled embedding to the 3D coordinate estimate.

Before training, each signal is downsampled by a factor of \num{20} 
and normalized by the maximum column norm $\max_{i=1,\ldots,N} \| \pmb{S}_i \|_2$, where $\pmb{S}_i$ is the $i$-th column of $\pmb{S}$. The target positions were scaled by a factor of \num{5}.
The measured noise samples are scaled by their standard deviation and 
added dynamically during training with a random scale factor in the range \numrange{0.0}{0.1}.
The model is trained for \num{100} epochs using the AdamW optimizer~\cite{loshchilov2018decoupled} 
with weight decay of \num{1E-4} and learning rate of \num{1E-4}, reduced by a factor of \num{0.5} 
after 32K, 64K, 96K, 128K, and 160K iterations.

\subsection{Experiment}

\begin{figure*}[tp]
    \centering
    \includegraphics[width=\textwidth]{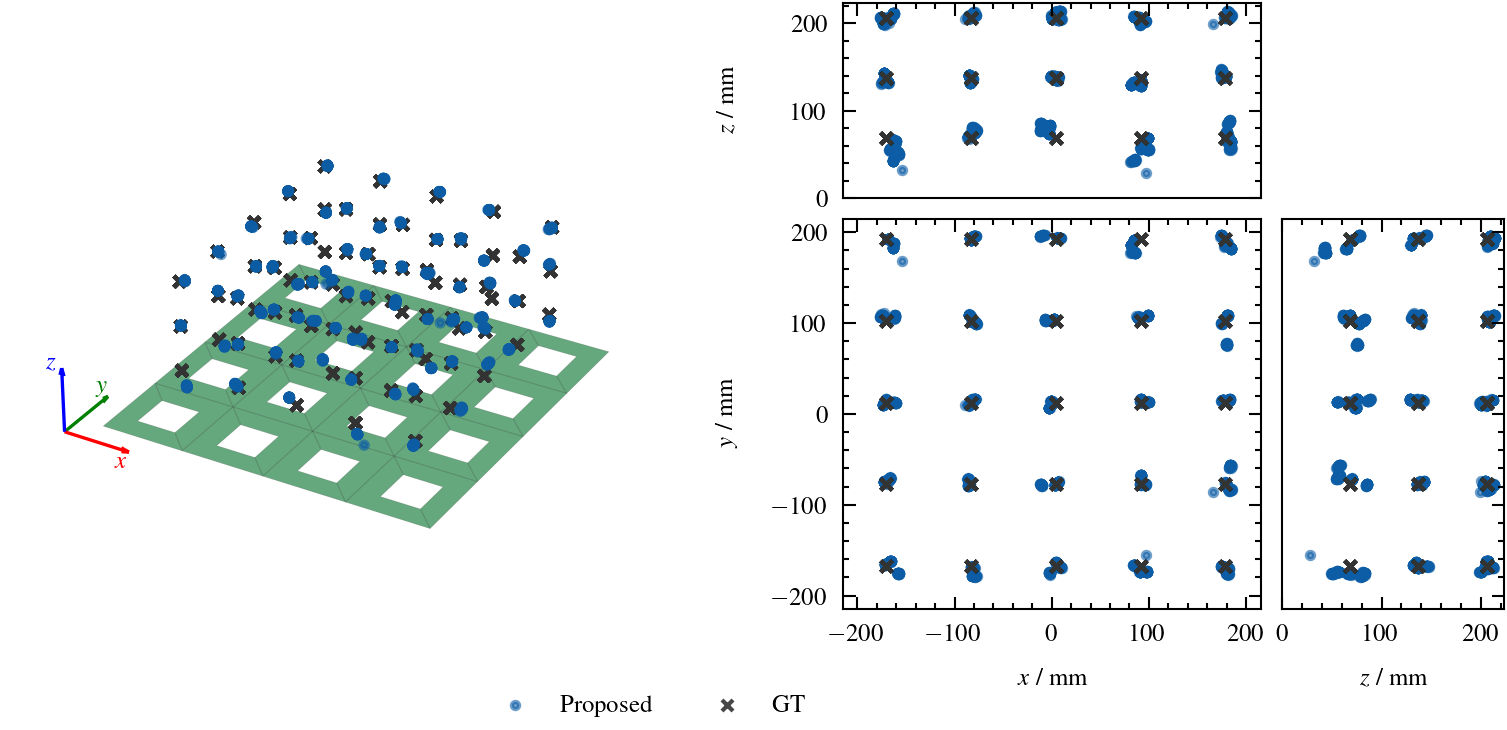}
    \caption{
        Estimated and \ac{GT} positions of the \ac{MMR} for each frame and grid point acquired in the experiment.
        The left panel shows the 3D spatial distribution together with the receive-coil geometry, while the right panels show the corresponding orthogonal projections 
        ($x$-$y$, $x$-$z$, and $z$-$y$). All coordinates are reported in millimeters.
    }
    \label{fig:results}
\end{figure*}

For the experimental validation of the proposed approach,
we use an \ac{MMR} composed of two cylindrical (P/N 20451,
MAGSY GmbH, Germany) neodymium–iron–boron
magnets with uniform diametral magnetization (\SI{4}{mm} diameter, \SI{4}{mm} height, separated by an approximately \SI{2}{mm} gap) 
mounted in a cuboidal 3D-printed housing (\SI{11}{mm} width, approximately \SI{30}{mm} height)~\cite{11139087}.

The coil array consists of \num{16} rectangular coils placed on a $4 \times 4$ grid with no margins.
Each coil has an outer dimension of \SI{122}{mm} and \num{60} windings spanning \SI{32}{mm} toward the center of the coil.
The coils are used for both signal transmission and reception via a \ac{TxRx} circuit. 
A double-pole double-throw electromagnetic relay is used as the \ac{TxRx} switch~\cite{11330979}. 
The received signal is passed through a low-noise amplifier based on
the ADA4898-2 operational amplifier (Analog Devices Inc., USA). 
The combined DAC/ADC system RedPitaya STEMlab 125-14 (Red Pitaya, Slovenia) 
is employed for signal generation and data acquisition~\cite{Hackelberg2022}.

The \ac{MMR} is moved by a three-axis robot with \SI{\pm0.025}{\mm} repeatability (Stepcraft M.500, Stepcraft GmbH \& Co. KG, Menden, Germany) 
on a \numproduct{5 x 5 x 3} grid with \SI{87.25}{mm}, \SI{90}{mm}, and \SI{69}{mm} spacing 
in the $x$, $y$, and $z$ directions, respectively. 
At each grid position, we perform a static measurement, i.e., without moving the \ac{MMR}, and acquire \num{25} frames, each with a \SI{0.88}{s} readout sampled at \SI{61035}{Hz}.
The orientation of the \ac{MMR} remains fixed during the experiment.

\subsection{Evaluation}

The acquired data are preprocessed by downsampling and scaling as described in the training section. 
The trained model is then applied to the preprocessed signals to obtain position estimates for each frame. 
We evaluate localization performance by computing the Euclidean distance between the estimated positions and the \ac{GT} positions recorded by the robot across all frames and grid points.
Because the predictions contain outliers, we report the 50th and 95th percentiles of the error distribution. 
\Acp{CI} for these quantiles are estimated using a percentile bootstrap. We benchmark the inference time for a single batch on CPU (AMD EPYC 9124) by repeating the model call \num{1000} times after \num{10} warm-up repetitions and report the minimum time over runs~\cite{chen_robust_2016}.

\section{Results}

The inferred positions of the \ac{MMR} for each frame, together with the \ac{GT} positions, are shown in Fig.~\ref{fig:results}. 
The proposed method correctly recovers the grid structure of the measured positions. 
The median Euclidean distance error across all frames and grid points is \SI{7.26}{mm} (95\% \ac{CI}: \SI{7.12}{mm}--\SI{7.46}{mm}), 
while the 95th percentile is \SI{23.7}{mm} (95\% \ac{CI}: \SI{23.0}{mm}--\SI{24.6}{mm}). 
We observe greater variability closer to the coil surface, where the error is dominated by uncertainty along the $z$ direction.
The minimum inference time is \SI{13}{\ms} for a single batch.

\section{Discussion and Conclusion}

The accuracy of the proposed method is comparable to the size of the \ac{MMR} sensor itself and 
corresponds to 6\% of a single coil size, 
demonstrating the feasibility of an end-to-end deep learning approach for \ac{MMR} localization. 
In addition, inference requires only \SI{13}{\ms} per batch, which is a small fraction of the \SI{0.88}{\s} frame duration, making the method suitable for real-time applications. 
The increased variability near the coil surface can be attributed to the normalization of the input signals. At these positions, the signal is dominated by a single coil rather than by the relative signal strengths across the array, and the normalization therefore reduces the information available to the model.

In this work, we focused on localization. In principle, however, the proposed approach can be extended to estimate additional parameters of the signal model, such as orientation and sensing-related quantities. This extension may require either separate models for individual parameters or a single model with multiple output heads and a multi-objective loss function. A key assumption of the proposed approach is that simulated data can adequately capture the distribution of real measurements. More accurate modeling of the system dynamics and noise could further improve the method. Future work may also investigate alternative normalization strategies and different model architectures.

\bibliographystyle{IEEEtran}

\bibliography{ref}

\end{document}